\documentclass[conference,a4paper]{IEEEtran}

\usepackage{amsmath}
\usepackage{amssymb}
\usepackage[T1]{fontenc}
\usepackage[utf8]{inputenc}
\usepackage{url}

\usepackage{setspace}

\newcommand{\abs}[1]{\left| #1 \right|}

\newcommand{\okra}[1]{\left( #1 \right)}
\newcommand{\kwad}[1]{\left[ #1 \right]}
\newcommand{\klam}[1]{\left\{ #1 \right\}}

\newcommand{\floor}[1]{\left\lfloor #1 \right\rfloor}

\DeclareMathOperator{\sred}{\mathbf{E}}
\newcommand{\boole}[1]{{\bf 1}{\klam{#1}}}

\newtheorem{theorem}{Theorem}

\begin{document}

\author{\IEEEauthorblockN{{\L}ukasz D\k{e}bowski}
  \IEEEauthorblockA{Institute of Computer Science\\ Polish Academy of
    Sciences\\ 01-248 Warszawa, Poland \\ Email:
    ldebowsk@ipipan.waw.pl}}

\title{Consistency of the Plug-In Estimator of the Entropy Rate for
  Ergodic Processes} \date{}

\pagestyle{empty}   
\maketitle

\begin{abstract}
  A plug-in estimator of entropy is the entropy of the distribution
  where probabilities of symbols or blocks have been replaced with
  their relative frequencies in the sample. Consistency and asymptotic
  unbiasedness of the plug-in estimator can be easily demonstrated in
  the IID case. In this paper, we ask whether the plug-in estimator
  can be used for consistent estimation of the entropy rate $h$ of a
  stationary ergodic process. The answer is positive if, to estimate
  block entropy of order $k$, we use a sample longer than
  $2^{k(h+\epsilon)}$, whereas it is negative if we use a sample
  shorter than $2^{k(h-\epsilon)}$. In particular, if we do not know
  the entropy rate $h$, it is sufficient to use a sample of length
  $(\abs{\mathbb{X}}+\epsilon)^{k}$ where $\abs{\mathbb{X}}$ is the
  alphabet size. The result is derived using $k$-block coding. As a
  by-product of our technique, we also show that the block entropy of
  a stationary process is bounded above by a nonlinear function of the
  average block entropy of its ergodic components. This inequality can
  be used for an alternative proof of the known fact that the entropy
  rate a stationary process equals the average entropy rate of its
  ergodic components.
%
\end{abstract}




\section{Results}
\label{secIntroduction}

Nonparametric entropy estimation is a task that requires a very large
amount of data. This problem has been studied mostly in the IID case,
see a review of literature in \cite{JiaoOthers15}. Moreover, the novel
results of \cite{JiaoOthers15} state that it is impossible to estimate
entropy of a distribution with a support size $S$ using an IID sample
shorter than of order $S/\log S$, whereas it is possible for a sample
longer than of order $S/\log S$, and a practical estimator achieving
this bound has been exhibited. Earlier, in \cite{Zhang12}, another
entropy estimator was proposed which, for a finite alphabet, has a
bias exponentially decreasing with the sample length. The exponential
decay of the bias is, however, too slow to beat the $S/\log S$ sample
bound.

In this paper we would like to pursue the more difficult and less
recognized question of entropy estimation for general stationary
ergodic processes, cf.\ \cite{Debowski16}.  For a stationary process
$(X_i)_{i=-\infty}^{\infty}$ over a finite alphabet $\mathbb{X}$,
consider the blocks of consecutive random symbols $X_k^l=(X_i)_{k\le
  i\le l}$. Consider then the true block distribution
\begin{align}
 p_k(w)=P(X_{i+1}^{i+k}=w) 
\end{align}
and the empirical distribution
\begin{align}
  p_k(w,X_1^n)&=
  \frac{1}{\floor{n/k}}\sum_{i=1}^{\floor{n/k}} \boole{X_{i(k-1)+1}^{ik}=w}
  .
\end{align}
Having denoted the entropy of a discrete distribution
\begin{align}
  H(p)=-\sum_{w:p(w)>0} p(w)\log p(w)
  ,
\end{align}
let the block entropy be
\begin{align}
 H(k)=H(p_k)
\end{align}
with the associated entropy rate
\begin{align}
  h=\lim_{n\rightarrow\infty} H(k)/k
  .
\end{align}
As shown in \cite{MartonShields94}, for the variational distance
\begin{align}
  \abs{p-q}:=\sum_{w} \abs{p(w)-q(w)}
  ,
\end{align}
we have
\begin{align}
 \lim_{k\rightarrow\infty} \abs{p_k-p_k(\cdot,X_1^{n(k)})}=0, 
\end{align}
if we put $n(k)\ge 2^{k(h+\epsilon)}$ for IID processes as well as for
irreducible Markov chains, for functions of irreducible Markov chains,
for $\psi$-mixing processes, and for weak Bernoulli processes.  This
result suggests that sample size $n(k)\ge 2^{k(h+\epsilon)}$ may be
sufficient for estimation of block entropy $H(k)$.

Let us state our problem formally. The plug-in estimator of the block
entropy is
\begin{align}
  H(k,X_1^n)&=H(p_k(\,\cdot\,,X_1^n))
  ,
\end{align}
as considered e.g.\ by \cite{ZhangZhang12}. 
Since 
\begin{align}
  \sred p_k(w,X_1^n)=P(X_{i+1}^{i+k}=w)
\end{align}
then, applying the Jensen inequality, we obtain
\begin{align}
  \label{ExpPlugIn}
  \sred H(k,X_1^n)&\le H(k)
\end{align}
so the plug-in estimator is a biased estimator of $H(k)$. The bias of
the plug-in estimator can be quite large since by inequality
$p_k(w,X_1^n)\ge \floor{n/k}^{-1}$ for $p_k(w,X_1^n)>0$ we also have
\begin{align}
  \label{ASPlugIn}
  H(k,X_1^n)&\le\log \floor{n/k}
  .
\end{align}

The plug-in estimator $H(k,X_1^n)$ depends on two arguments: the block
length $k$ and the sample $X_1^n$. If we fix the block length $k$ and
let the sample size $n$ tend to infinity, we obtain a consistent and
asymptotically unbiased estimator of the block entropy $H(k)$. Namely,
for a stationary ergodic process,
\begin{align}
 \lim_{n\rightarrow\infty} H(k,X_1^n)=H(k) \text{ a.s.} 
\end{align}
 by the
ergodic theorem and hence
\begin{align}
 \lim_{n\rightarrow\infty} \sred H(k,X_1^n)=H(k) 
\end{align}
by inequality (\ref{ExpPlugIn}) and the Fatou lemma. These results
generalize what is known for the IID case \cite{ZhangZhang12}.

Now the question arises what $n(k)$ we should choose so that
$H(k,X_1^{n(k)})/k$ be a consistent estimator of the entropy
rate. Using a technique based on source coding, which is different
than used in \cite{MartonShields94}, we may establish some positive
result in a more general case than considered in
\cite{MartonShields94}:
\begin{theorem}
  \label{theoPlugIn}
  Let $(X_i)_{i=-\infty}^{\infty}$ be a stationary ergodic process
  over a finite alphabet $\mathbb{X}$.  For any $\epsilon>0$ and
  \begin{align}
    \label{GoodN}
    n(k)\ge 2^{k(h+\epsilon)}
    ,
  \end{align}
  we have
  \begin{align}
    \label{PlugInExp}
    \lim_{k\rightarrow\infty} \sred H(k,X_1^{n(k)})/k&=h
                                                    ,
    \\
    \label{PlugInAS}
    \liminf_{k\rightarrow\infty} H(k,X_1^{n(k)})/k&= h \text{ a.s.}
                                                    ,
                                                    \\
    \label{PlugInProb}
    \forall_{\eta>0} \lim_{k\rightarrow\infty} P\okra{H(k,X_1^{n(k)})/k-h>\eta}&= 0
                                                       .
  \end{align}
\end{theorem}

According to Theorem \ref{theoPlugIn}, for the sample size
(\ref{GoodN}) the plug-in estimator $H(k,X_1^{n(k)})/k$ of the entropy
rate $h$ is consistent in probability.  In contrast, applying
inequality (\ref{ASPlugIn}) for $\epsilon>0$ and
\begin{align}
  \label{BadN}
  n(k)\le 2^{k(h-\epsilon)}
\end{align}
yields
\begin{align}
  \limsup_{k\rightarrow\infty} H(k,X_1^{n(k)})/k\le h-\epsilon \text{ a.s.}
\end{align}
Hence the sample size (\ref{BadN}) is insufficient to
obtain a consistent estimate of the entropy rate $h$ using the plug-in
estimator.

Let us observe that in general there are two different kinds of random
entropy bounds for stationary processes:
\begin{enumerate}
\item Random upper bounds $K(X_1^n)$ based on universal coding
  \cite{ZivLempel77,NeuhoffShields98} or universal prediction
  \cite{Ryabko10}: For these bounds, we have Kraft inequality
  $\sum_{x_1^n}2^{-K(x_1^n)}\le 1$.  Therefore, for a stationary
  process, we have the source coding inequality $\sred K(X_1^n)\ge
  H(n)$ and the Barron inequality
  \begin{align}
    \label{BarronII}
    P\okra{K(X_1^{n})+\log P(X_1^{n})\le -m}\le 2^{-m}
  \end{align}
  \cite[Theorem 3.1]{Barron85b}. Moreover, for a stationary ergodic
  process, we have
  \begin{align}
    \label{ConsistencyPlus}
    \lim_{n\rightarrow\infty}  \sred K(X_1^n)/n&=h
    ,
    \\
    \lim_{n\rightarrow\infty}  K(X_1^n)/n&=h \text{ a.s.}
  \end{align}
  In particular, these conditions hold for 
  \begin{align}
    K(X_1^n)=\min_k K(k,X_1^n)
    ,
  \end{align}
  where $K(k,X_1^n)$ is the length of the code which will
  be considered for proving Theorem \ref{theoPlugIn} in Section
  \ref{secProof}, cf. \cite{NeuhoffShields98}.
\item Random lower bounds, such as the plug-in estimator
  $H(k,X_1^n)$: As we have seen, for a stationary process, we
  have (\ref{ExpPlugIn}), whereas for a stationary ergodic process, we
  have (\ref{PlugInExp})--(\ref{PlugInProb}).
\end{enumerate}
Both quantities $K(X_1^n)$ and $H(k,X_1^n)$ can be used for
estimation of the entropy rate $h$.

When applying $H(k,X_1^{n(k)})$ for the estimation of entropy rate, we
are supposed not to know the exact value of $h$. Therefore, the choice
of minimal admissible $n(k)$ is not so trivial.  According to Theorem
\ref{theoPlugIn}, we may put
$n(k)=(\abs{\mathbb{X}}+\epsilon)^k$. This bound is, however,
pessimistic, especially for processes with a vanishing entropy rate
$h=0$, cf. \cite{Debowski15g}. Having a random upper bound of the
block entropy $K(X_1^k)$, we may also put $n(k)=2^{\sred
  K(X_1^k)+k\epsilon}$.

A question arises whether we can improve Theorem
\ref{theoPlugIn}. Thus, let us state three open problems:
\begin{enumerate}
\item Does the equality
  \begin{align}
    \lim_{k\rightarrow\infty} H(k,X_1^{n(k)})/k=h \text{ a.s.}
  \end{align}
  hold true in some cases? In other words, is the plug-in estimator
  $H(k,X_1^{n(k)})/k$ an almost surely consistent estimator of the
  entropy rate?
\item What happens for $\lim_{k\rightarrow\infty} k^{-1}\log n(k)=h$?
  In particular, can Theorem \ref{theoPlugIn} be strengthened by
  setting $n(k)$ equal to some random stopping time, such as
\begin{align}
  \label{PossibleN}
  n(k)=2^{K(X_1^k)}
  ,
\end{align}
where $K(X_1^k)$ is a length of a universal code for $X_1^k$?
\item The plug-in estimator is not optimal in the IID case
  \cite{JiaoOthers15}. Can we propose a better estimator of the
  entropy rate also for an arbitrary stationary ergodic process?
\end{enumerate}

Another class of less clearly stated problems concerns comparing the
entropy estimates $K(X_1^k)$ and $H(k,X_1^{n(k)})$.  Although the gap
between these estimates is closing when divided by $k$, i.e.,
\begin{align}
  \lim_{k\rightarrow\infty}  \sred \kwad{K(X_1^k)-H(k,X_1^{n(k)})}/k=0
  ,
\end{align}
the difference $K(X_1^k)-H(k,X_1^{n(k)})$ can be arbitrarily large. To
see it, let us note that inequalities $\sred K(X_1^k)\ge H(k)$ and
$H(k)\ge \sred H(k,X_1^{n})$ hold for any stationary process,
regardless whether it is ergodic or not. Hence, by the ergodic
decomposition \cite{GrayDavisson74b}, we have $\sred K(X_1^k)\ge
H(X_1^k)$ and $H(X_1^k|\mathcal{I})\ge \sred H(k,X_1^{n})$, where
$\mathcal{I}$ is the shift-invariant algebra of a stationary process
$(X_i)_{i=-\infty}^{\infty}$, $H(X_1^k)=H(k)$ is the entropy of
$X_1^k$, and $H(X_1^k|\mathcal{I})$ is the conditional entropy of
$X_1^k$ given $\mathcal{I}$. Consequently,
\begin{align}
  \sred\kwad{K(X_1^k)-H(k,X_1^n)} &\ge
  H(X_1^k)-H(X_1^k|\mathcal{I})
  \nonumber\\
  &=I(X_1^k;\mathcal{I}),
\end{align}
where $I(X_1^k;\mathcal{I})$ is the mutual information between block
$X_1^k$ and the shift-invariant algebra $\mathcal{I}$. In fact, for an
arbitrary stationary process, the mutual information
$I(X_1^k;\mathcal{I})$ can grow as fast as any sublinear function,
cf.\ \cite{Debowski12,Debowski15g}.

Whereas there is no universal sublinear upper bound for mutual
information $I(X_1^k;\mathcal{I})=H(X_1^k)-H(X_1^k|\mathcal{I})$, we
may ask whether there is an upper bound for entropy $H(X_1^{n})$ in
terms of a function of conditional entropy $H(X_1^k|\mathcal{I})$ for
an arbitrary stationary process and $n\ge k$. Using the code from the
proof of Theorem \ref{theoPlugIn}, we can provide this bound:
\begin{theorem}
  \label{theoRateDifference}
  For a stationary process $(X_i)_{i=-\infty}^{\infty}$, natural
  numbers $p$ and $k$, $n=pk$, and a real number $m\ge 1$,
  \begin{align}
    &\frac{H(X_1^{n})}{n}-\frac{H(X_1^k|\mathcal{I})}{k} \le 
    \frac{2}{k}+
    \frac{2}{n}\log k +
    3\log\abs{\mathbb{X}}\times
    \nonumber\\
    \label{RateDifferenceBound}
    &\times\okra{
      \frac{1}{m} + 
      \okra{1-\frac{1}{m}}
      \sigma\okra{mH(X_1^k|\mathcal{I})-\log\frac{n}{k}}+
      \frac{k}{n}}
    ,
  \end{align}
  where $\sigma(y)=\min(\exp(y),1)$.
\end{theorem}

Theorem \ref{theoRateDifference} states that the block entropy of a
stationary process is bounded above by a nonlinear function of the
average block entropy of its ergodic components. We suppose that this
inequality can be strengthened if there exists a better estimator of
the block entropy than the plug-in estimator.  A simple corollary of
Theorem \ref{theoRateDifference} is that 
\begin{align}
  \label{RateEquality}
  \lim_{k\rightarrow\infty} H(X_1^k|\mathcal{I})/k=h
  ,
\end{align}
a fact usually proved by the ergodic decomposition \cite[Theorem
5.1]{GrayDavisson74b}. To derive (\ref{RateEquality}) from
(\ref{RateDifferenceBound}), we first put $n\rightarrow\infty$ and
next $m\rightarrow\infty$ and $k\rightarrow\infty$.

In the following, in Section \ref{secProof} we prove Theorem
\ref{theoPlugIn}, whereas in Section \ref{secProofII} we prove
Theorem \ref{theoRateDifference}.

\section{Proof of Theorem \ref{theoPlugIn}}
\label{secProof}

Our proof of Theorem \ref{theoPlugIn} applies source coding. To be
precise, it rests on a modification of the simplistic universal code
by Neuhoff and Shields \cite{NeuhoffShields98}. The Neuhoff-Shields
code is basically a $k$-block code with parameter $k$ depending on the
string $X_1^n$. In the following, we will show that the plug-in
estimator $H(k,X_1^n)$ multiplied by $n/k$ is the dominating term in
the length of a modified $k$-block code for $X_1^n$ by the results of
\cite{NeuhoffShields98,OrnsteinWeiss90}. This length cannot be shorter
than $nh$ so the expectation of $H(k,X_1^n)/k$ must tend to $h$.

The idea of a $k$-block code is that we first describe a code book,
i.e., we enumerate the collection of blocks $w$ of length $k$
contained in the compressed string $X_1^n$ and their frequencies
$np_k(w,X_1^n)$, and then we apply the Shannon-Fano coding to $X_1^n$
partitioned into blocks from the code book.  Let $D(k,X_1^n)$ be the
number of distinct blocks of length $k$ contained in the compressed
string $X_1^n$. Formally,
\begin{align}
  D(k,X_1^n)=
  \abs{\klam{w\in\mathbb{X}^k:\exists_{i\in{1,...,\floor{n/k}}}
  X_{(i-1)k+1}^{ik}=w}}
  .
\end{align}
To fully describe $X_1^n$ in terms of a $k$-block code we have to
specify, cf. \cite{NeuhoffShields98}:
\begin{enumerate}
\item what $k$ is (description length $2\log k$),
\item what $D(k,X_1^n)$ is (description length $\log\floor{n/k}$),
\item what the code book is (we have to specify the Shannon-Fano code
  word for each $k$-block, hence the description length is $\le
  (k\log\abs{\mathbb{X}}+2\log\floor{n/k}) D(k,X_1^n)$),
\item what the Shannon-Fano code words for block $X_1^{k\floor{n/k}}$
  are (description length $\le\floor{n/k}(H(k,X_1^n)+1)$),
\item what the remaining block $X_{k\floor{n/k}+1}^n$ is (description
  length $\le k\log\abs{\mathbb{X}}$).
\end{enumerate}
Hence quantity 
\begin{align}
  &2\log k+\frac{n}{k}\okra{H(k,X_1^n)+1}+
  \nonumber\\
  &\phantom{=}+\okra{k\log\abs{\mathbb{X}}+2\log\frac{n}{k}} \okra{D(k,X_1^n)+1}
\end{align}
is an upper bound for the length of the $k$-block code.

For our application, the $k$-block code has a deficiency that very
rare blocks have too long codewords, which leads to an unwanted
explosion of term $2\log\frac{n}{k}$ in the upper bound of the code
length for $n\rightarrow\infty$. Hence let us modify the $k$-block
code so that a $k$-block is Shannon-Fano coded if and only if its
Shannon-Fano code word is shorter than $k\log\abs{\mathbb{X}}$,
whereas it is left uncoded otherwise. In the coded sequence, to
distinguish between these two cases, we have to add some flag, say $0$
before the Shannon-Fano code word and $1$ before the uncoded block.
In this way, to fully describe $X_1^n$ in terms of the modified
$k$-block code we have to specify:
\begin{enumerate}
\item what $k$ is (description length $2\log k$),
\item what the number of used distinct Shannon-Fano code words is
  (description length $k\log\abs{\mathbb{X}}$),
\item what the code book is (we have to specify the Shannon-Fano code
  word for each coded $k$-block, hence the description length is $\le
  3k\log\abs{\mathbb{X}} D(k,X_1^n)$),
\item what the sequence of code words for block $X_1^{k\floor{n/k}}$
  is (description length $\le\floor{n/k}(H(k,X_1^n)+2)$),
\item what the remaining block $X_{k\floor{n/k}+1}^n$ is (description
  length $\le k\log\abs{\mathbb{X}}$).
\end{enumerate}
In view of this, quantity
\begin{align}
  K(k,X_1^n)&=2\log k+\frac{n}{k}\okra{H(k,X_1^n)+2}+
  \nonumber\\
  &\phantom{=}+3k\log\abs{\mathbb{X}} \okra{D(k,X_1^n)+1}
\end{align}
is an upper bound for the length of the modified $k$-block code.

Since the $k$-block code is an instantaneous code, the upper bound for
its length satisfies Kraft inequality
$\sum_{k,x_1^n}2^{-K(k,x_1^n)}\le 1$.  Therefore, we have $\sred
K(k,X_1^n)\ge H(n)$, whereas by the Barron inequality
\begin{align}
  \label{Barron}
  P\okra{K(k,X_1^{n})+\log P(X_1^{n})\le -m}\le 2^{-m}
\end{align}
\cite[Theorem 3.1]{Barron85b}, the Borel-Cantelli lemma, and the
Shannon-McMillan-Breiman theorem
\begin{align}
  \label{SMB}
  \lim_{n\rightarrow\infty} \frac{\kwad{-\log P(X_1^{n})}}{n} = h
  \text{ a.s.}
\end{align}
\cite{AlgoetCover88}, we obtain
\begin{align}
  \liminf_{k\rightarrow\infty} \frac{K(k,X_1^{n(k)})}{n(k)}
  \ge
  \lim_{k\rightarrow\infty} \frac{\kwad{-\log P(X_1^{n(k)})}}{n(k)}
  =
  h
  \text{ a.s.}
\end{align}

According to \cite[Theorem 2]{OrnsteinWeiss90},
for each $\delta>0$ almost surely there exists $k_0$ such that for
all $k\ge k_0$ and $n>2^{kh}$ we have
\begin{align}
  D(k,X_1^n)\le 2^{k(h+\delta)}+\frac{n}{k}\delta
  .
\end{align}
Hence for $n(k)\ge 2^{k(h+\epsilon)}$ and
$\delta<\epsilon$, we have almost surely
\begin{align}
  h&\le \liminf_{k\rightarrow\infty} \frac{K(k,X_1^{n(k)})}{n(k)}
  \nonumber\\
   &=\liminf_{k\rightarrow\infty} 
     \left(\frac{3k}{n(k)}\log\abs{\mathbb{X}}D(k,X_1^{n(k)})
     +\frac{H(k,X_1^{n(k)})}{k}\right)
  \nonumber\\
   &\le\liminf_{k\rightarrow\infty} 
     \left(3\log\abs{\mathbb{X}}
     \okra{k2^{-k(\epsilon-\delta)}
     +\delta}+\frac{H(k,X_1^{n(k)})}{k}\right)
  \nonumber\\
   &=3\log\abs{\mathbb{X}}
     \delta+\liminf_{k\rightarrow\infty} \frac{H(k,X_1^{n(k)})}{k}
     .
\end{align}
Since $\delta$ can be chosen arbitrarily small then
\begin{align}
  \label{LimInf}
  \liminf_{k\rightarrow\infty}  \frac{H(k,X_1^{n(k)})}{k}\ge h 
  \text{ a.s.}
\end{align}

In contrast, inequality (\ref{ExpPlugIn}) implies
\begin{align}
  \limsup_{k\rightarrow\infty} \frac{\sred H(k,X_1^{n(k)})}{k}\le h
  .
\end{align}
Hence, by the Fatou lemma and inequality (\ref{LimInf}), we have
\begin{align}
  \label{ELimInf}
  h=\sred \liminf_{k\rightarrow\infty} \frac{H(k,X_1^{n(k)})}{k}
  =\lim_{k\rightarrow\infty} \frac{\sred H(k,X_1^{n(k)})}{k}
  ,
\end{align}
i.e., equality (\ref{PlugInExp}) is established. By inequality
(\ref{LimInf}) and equality (\ref{ELimInf}), we also obtain equality 
(\ref{PlugInAS}).

The proof of statement (\ref{PlugInProb}) requires a few additional
steps. Denoting $X^+=X\boole{X>0}$ and $X^-=-X\boole{X<0}$, we obtain
from Markov inequality, inequality (\ref{ExpPlugIn}), and inequality
$(X+Y)^-\le X^-+Y^-$ that
 \begin{align}
   &\eta P\okra{\frac{H(k,X_1^{n(k)})}{k}-h>\eta}\le
   \sred\kwad{\frac{H(k,X_1^{n(k)})}{k}-h}^+
   \nonumber
   \\
   &=
   \sred\kwad{\frac{H(k,X_1^{n(k)})}{k}-h}+
   \sred\kwad{\frac{H(k,X_1^{n(k)})}{k}-h}^-
   \nonumber
    \\
   &\le
   \kwad{\frac{H(k)}{k}-h}+
   \sred\kwad{\frac{H(k,X_1^{n(k)})}{k}-\frac{K(k,X_1^{n(k)})}{n(k)}}^-+
   \nonumber
   \\
   &\phantom{=}+
   \sred\kwad{\frac{K(k,X_1^{n(k)})}{n(k)}+\frac{\log P(X_1^{n(k)})}{n(k)}}^-+
   \nonumber
   \\
   &\phantom{=}+
   \sred\kwad{\frac{\kwad{-\log P(X_1^{n(k)})}}{n(k)}-h}^-
   \label{FourTerms}
   .
 \end{align}

 Now we will show that all four terms on the RHS of (\ref{FourTerms})
 tend to $0$, which is sufficient to establish
 (\ref{PlugInProb}). First,
 \begin{align}
   \lim_{k\rightarrow\infty} \kwad{\frac{H(k)}{k}-h}=0
 \end{align}
 by the definition of the entropy rate. Second, 
 \begin{align}
   \lim_{k\rightarrow\infty} 
   \sred\kwad{\frac{H(k,X_1^{n(k)})}{k}-\frac{K(k,X_1^{n(k)})}{n(k)}}^-=0
 \end{align}
 since
 \begin{align}
   \lim_{k\rightarrow\infty} \frac{k}{n(k)}\sred
   D(k,X_1^{n(k)})=0
 \end{align}
 by the result of \cite[Eq. (8)]{NeuhoffShields98}. Third,
 \begin{align}
   \lim_{n\rightarrow\infty} 
   \frac{1}{n}\sred\kwad{K(k,X_1^{n})+\log P(X_1^{n})}^-=0
 \end{align}
 since
 \begin{align}
   \sred\kwad{K(k,X_1^{n})+\log P(X_1^{n})}^-\le \sum_{m=0}^\infty m
   2^{-m}<\infty
 \end{align}
 by the Barron inequality (\ref{Barron}). Fourth,
 \begin{align}
   \lim_{n\rightarrow\infty} 
   \sred\kwad{\frac{\kwad{-\log P(X_1^{n})}}{n}-h}^-=0
 \end{align}
 since the Shannon-McMillan-Breiman theorem (\ref{SMB}) implies
 convergence in probability, i.e., for all $\epsilon>0$, we have
 \begin{align}
   \lim_{n\rightarrow\infty}
   P\okra{\frac{\kwad{-\log P(X_1^{n})}}{n} - h < -\epsilon}=0
   .
 \end{align}
 Hence
 \begin{align}
   &\sred\kwad{\frac{\kwad{-\log P(X_1^{n})}}{n}-h}^-
   \nonumber
   \\
   &\le h P\okra{\frac{\kwad{-\log P(X_1^{n})}}{n} - h < -\epsilon}
   \nonumber
   \\
   &\phantom{=}
   + \epsilon P\okra{\frac{\kwad{-\log P(X_1^{n})}}{n} - h \ge -\epsilon}
 \end{align}
 tends to a value smaller than $\epsilon$, where $\epsilon$ is
 arbitrarily small.

\section{Proof of Theorem \ref{theoRateDifference}}
\label{secProofII}

For the code from the proof of Theorem \ref{theoPlugIn}, we have
\begin{align}
  \label{RateDifference}
  &\frac{H(X_1^{n})}{n}-\frac{H(X_1^k|\mathcal{I})}{k} \le
  \sred\kwad{\frac{K(k,X_1^{n})}{n}-\frac{H(k,X_1^{n})}{k}}
  \nonumber\\
  &=\frac{2}{k}+\frac{2}{n}\log k+\frac{3k}{n}\log\abs{\mathbb{X}}
  \okra{\sred D(k,X_1^n)+1}
  .
\end{align}
Then, following the idea of \cite[Theorem 1]{Debowski16}, we may
express the number of distinct $k$-blocks as
\begin{align}
  D(k,X_1^n)=\sum_{w\in\mathbb{X}^k}\boole{\sum_{i=1}^{n/k}
    \boole{X_{(i-1)k+1}^{ik}=w}\ge 1}.
\end{align}
Hence by the Markov inequality,
\begin{align}
  &\sred\okra{D(k,X_1^n)|\mathcal{I}}
  \nonumber\\
  &= \sum_{w\in\mathbb{X}^k} P\okra{\sum_{i=1}^{n/k}
    \boole{X_{(i-1)k+1}^{ik}=w}\ge 1\middle|\mathcal{I}}
  \nonumber\\
  &\le \sum_{w\in\mathbb{X}^k} \min\kwad{1,\sred\okra{\sum_{i=1}^{n/k}
      \boole{X_{(i-1)k+1}^{i+k}=w}\middle|\mathcal{I}}}
  \nonumber\\
  &= \sum_{w\in\mathbb{X}^k}
  \min\kwad{1,\frac{n}{k}P(X_1^k=w|\mathcal{I})}
  \nonumber\\
  &=
  \frac{n}{k}\sred\okra{\min\okra{\kwad{\frac{n}{k}P(X_1^k|\mathcal{I})}^{-1},1}
    \middle|\mathcal{I}}
  .
\end{align}
In consequence,
\begin{align}
  \label{DictionarySigma}
  \frac{k}{n}\sred D(k,X_1^n)\le \sred\sigma\okra{-\log
    P(X_1^k|\mathcal{I})-\log\frac{n}{k}}
  ,
\end{align}
where $\sred\kwad{-\log
  P(X_1^k|\mathcal{I})}=H(X_1^k|\mathcal{I})$. Therefore, using another
Markov inequality
\begin{align}
  P\okra{-\log P(X_1^k|\mathcal{I})\ge mH(X_1^k|\mathcal{I})}\le
  \frac{1}{m}
\end{align}
for $m\ge 1$, we further obtain from (\ref{DictionarySigma}) that
\begin{align}
  \label{DictionarySigmaII}
  \frac{k}{n}\sred D(k,X_1^n)\le  \frac{1}{m} + 
  \okra{1-\frac{1}{m}}\sigma\okra{mH(X_1^k|\mathcal{I})-\log\frac{n}{k}}
  .
\end{align}
Inserting (\ref{DictionarySigmaII}) into (\ref{RateDifference}) yields
the requested bound.

\bibliographystyle{IEEEtran}


\end{document}